\documentclass[12pt,titlepage]{article}

\usepackage{graphicx}

\usepackage[]{natbib}

\begin{document}



\title{Hydration of Krypton and Consideration of Clathrate
Models of Hydrophobic Effects from the Perspective of Quasi-Chemical Theory}



\author{Henry S. Ashbaugh, D. Asthagiri,
Lawrence  R.  Pratt\footnote{Theoretical Division, Los Alamos National
Laboratory, Los Alamos, NM 87545, USA},\\
 and \\
Susan B. 
Rempe\footnote{Sandia National
Laboratory, Albuquerque, NM 87185 USA}}



\date{\today}


\maketitle



\pagebreak

\begin{abstract}
{\em Ab initio\/} molecular dynamics (AIMD) results on a liquid
krypton-water system are presented and  compared to recent XAFS results
for the radial hydration structure for a Kr  atom in liquid water
solution. Though these AIMD calculations have important  limitations  of
scale, the comparisons with the liquid  solution results are
satisfactory and significantly different from the radial distributions
extracted from the data on the solid Kr/H$_2$O clathrate hydrate phase.
The calculations also produce the coordination number distribution that
can be examined for metastable coordination structures suggesting
possibilities for clathrate-like organization; none  are seen in these
results. Clathrate pictures of hydrophobic hydration are discussed, as
is the quasi-chemical theory  that should provide a basis for clathrate
pictures. Outer  shell contributions are discussed and accurately
estimated; they are positive and larger than the positive experimental
hydration free energy of Kr(aq), implying that inner shell contributions
must be negative and of comparable size. Clathrate-like inner shell
hydration structures on a Kr atom solute are obtained for some, but not
all, of the coordination number cases observed in the simulation.   The
structures found have a delicate stability; inner shell coordination
structures extracted from the simulation of the liquid, and then
subjected to quantum chemical optimization, always decomposed.
Interactions with the outer shell material are decisive in stabilizing
coordination structures observed in liquid solution and in clathrate
phases.   The  primitive quasi-chemical estimate that uses a 
dielectric model for the influence of the outer shell material on the
inner shell equilibria gives a contribution to hydration free energy that is
positive and larger than the experimental hydration free energy. The
``what are we to tell students'' question about hydrophobic hydration,
often answered with structural models such as clathrate pictures, is
then considered; we propose an alternative answer that is consistent
with successful molecular theories of hydrophobic  effects and based
upon distinctive observable properties of liquid water. Considerations
of parsimony, for instance Ockham's razor, then suggest that additional
structural hypotheses  in response to ``what are we to tell students''
aren't required at this stage.


\end{abstract}

\section{Introduction}

Kauzmann's analysis \cite{Kauzmann:59} established the topic of
hydrophobic effects as relevant to the structure, stability, and
function of soluble proteins.  Reminiscences on protein research of that
period emphasize that the  concept of hydrophobic stabilization of
globular proteins was non-trivial \cite{Tanford:97}. Decades have
passed, and the language of hydrophobic effects has become common. It
is, therefore, astonishing to note that consensus on a molecular scale
conceptualization of hydrophobic effects, and on molecular  theories,
has not been obtained.  For a recent example, see reference
\cite{LazaridisT:Solsvc}.  In the  present setting of  competing ideas,
theories, and selected results, it isn't uncommon to hear from
biophysical chemists, ``What are we to tell students?''  Of course, one
response is  to question what we tell ourselves.

In addressing the most primitive hydrophobic effects,  molecular theory
has achieved some surprising steps recently.  A compelling molecular
theory for primitive hydrophobic effects is now
available \cite{PrattLR:Molthe}, though a variety of more complex cases
haven't been similarly resolved \cite{PrattLR:CR02}.  Still,  the recent
theoretical advance means that we can scrutinize conflicting theories
more seriously and begin to build a more objective answer to the  ``what
are we to tell  students'' question.

The recent XAFS studies of the hydration of Kr in liquid water and
in a comparable solid Kr/H$_2$O clathrate phase \cite{BowronDT:Hydhfc} seem
particularly helpful to  the task of consolidation of molecular theories
of primitive hydrophobic effects.  These experiments were directed
toward characterization of the iceberg pictorial interpretations  and
conclusions of Frank \& Evans \cite{Frank:JCP:45} that were supported in part by the  known
structures of solid clathrate hydrates.

In the  panoply of {\em  pictures\/} of hydrophobic phenomena, clathrate
models are especially appropriate as initial targets for modern
scrutiny: it has  been  discovered recently how these structural models
might be developed logically as a basis for molecular theory of
hydrophobic effects \cite{HummerG:Newphe,Paulaitis:APC:02}. Those
approaches are {\em quasi-chemical\/} theories.  They  hold promise for
higher molecular resolution, and particularly for treating {\em context
hydrophobicity\/} in which  proximal hydrophilic groups affect the
hydration of hydrophobic moieties.

A drawback that can be foreseen for quasi-chemical theories is one that
is common to computational chemistry: several competing contributions
must be evaluated at an appropriate accuracy and assembled.  The net
hydration free energy of Kr(aq) is an order of magnitude smaller than
several of the primitive contributions to it. Every little thing counts
in the assembled results and experience in evaluating the various
required contributions is thin.  The chemically simple case of Kr(aq) is
an appropriate starting point because the various contributions that
must be evaluated seem to be in hand. Nevertheless, detailed
computational determination of the hydration free energy of Kr(aq) is 
not sought in the present effort, which instead takes the goals of
testing clathrate pictures of hydrophobic  hydration and of scoping
molecularly detailed quasi-chemical theories. 

The plan for this paper is as follows: in the next section we give some
background on clathrate models of hydrophobic hydration.  After that, we
discuss quasi-chemical theory in order to explain the coordination
number distribution that is the principal target of the following {\em
ab initio\/} molecular dynamics (AIMD) calculations. AIMD results for
the hydration of Kr(aq) are then compared with results from the XAFS
experiments.   To guide quasi-chemical descriptions of these phases, we
then estimate the inner and outer shell contributions required for the
most primitive approximations. The final sections discuss ``what are we
to  tell  students'' and identify conclusions.

Though AIMD calculations  might seem to be excessive for this chemically
simple problem, this approach is available and the results aren't
limited by assumptions of non-cooperativity of the interactions. Modern
work \cite{SWAMINATHANS:MONSSD,SWOPEWC:MOLMCS,WATANABEK:MOLSTH} on
hydrophobic hydration has suggested that attractive interactions between
solute and solvent water might play a role less trivial than
assumed by conventional van~der~Waals theories \cite{Chandler:83}. This point is
consistent with the fact that clathrate phases involving inert gases as
light as Ne are more exotic and are to be sought at relatively higher
pressures than similar phases involving heavier inert gases
\cite{DyadinYA:Clahhn,KosyakovVI:Simpet,zhao:02,Sloan}.

The present work is centered  on the physical chemical basics of a
foundational concept of molecular structural biology.  We hope this
focus follows the example provided by Kauzmann's wide-ranging
considerations of molecular science, including topics of water and
aqueous solutions \cite{Eisenberg} and molecular  biophysical chemistry.

\section{Clathrate Model Background}

The status of clathrate pictures of hydrophobicity is reflected
in the current literature;  for example, see
\cite{ChengYK:Surtdb}: ``Conventional views hold that the hydration
shell of small hydrophobic solutes is clathrate-like, characterized by
local cage-like hydrogen-bonding structures and a distinct loss in
entropy.'' An admirably clear and appropriately circumspect early
consideration of clathrate models for solutions of hydrophobic gases in
liquid water was given by Glew \cite{Glew:62}.  The data available at that
time suggested ``\ldots that the nature of the water solvent surrounding
weakly interacting aqueous solutes should be likened geometrically to
those co\"{o}rdination polyhedra experimentally observed in the solid
gas-hydrates.'' Still, ``in no sense is it considered that the water
molecules adjacent to the  solute are permanently immobilized or rigid
as in solid structures  \ldots'' \cite{Glew:62}.  The more recent
discussion of \cite{Dill:90:b} explains conventional views  of hydrophobic
hydration ``\ldots
that the organization of water molecules in the  first shell surrounding
the solute is like an `iceberg,' a clathrate, or a `flickering
cluster.'$\thinspace$'' It may be disconcerting that a series of
non-equivalent descriptors are used to describe the same phenomenon. This fact
reflects the imprecision of these conceptualizations and we won't
distinguish them further. Another much earlier expression of
such a physical view can be found in \cite{klotz}; at the same time a
clathrate model of liquid water {\em without} solutes was
proposed by Pauling \cite{Pauling:59}.  The review in \cite{SouthallNT:viethe}
discussed again Kauzmann's consideration \cite{Kauzmann:59} of
clathrate models of hydrophobic hydration, which argued that the degree
of ordering suggested by measured hydration entropy changes upon
dissolution was too small to be conceived as crystalline ordering.
Concluding, Kauzmann \cite{Kauzmann:59} stated: ``There  is no
justification for using the iceberg concept as a basis for the
hypothesis that protein molecules are surrounded and stabilized by
regions of ordered water molecules.''

On the solution structure side of this issue, Head-Gordon
\cite{HEADGORDONT:ISWAHG} examined the question of whether the 
structure of liquid water near hydrophobic inert gas molecules was
clathrate-like by exploiting computer simulation to compute the numbers
of H-bonded pentagons present near those solutes compared to bulk
regions  of liquids.  The idea was that pentagonal bonding structures
are evident in clathrate crystals so pentagons might be a suitable
diagnostic of clathrate-like structures in liquid solutions. Some
enhancement of the populations of pentagons was observed in the first
hydration shell of model CH$_4$ solute in liquid water, but the
conclusions were ultimately mixed and included that  ``it is important to
emphasize that no direct connection between structure and thermodynamics
is made --- {\em i.e.,\/} through a formal statistical mechanical
theory.''  

On the crystal side of this issue, clathrate phases with
impressively complex guests have been discovered
\cite{UdachinKA:comchs,UdachinKA:denech}. But it can  happen that,
though the water molecules are ordered, the structure of the guest
molecule is disordered.  That is opposite to the case sought in protein
crystallography. Earlier \cite{TEETERMM:WATHPA} analysis of crambin
crystals had identified a cluster of pentagonal rings of water molecules
at the surface of that protein. Those rings are extensively H-bonded to
the protein and crystal packing effects are
involved in principle.
It has been challenged \cite{LipscombLA:Clahap} whether pentagonal H-bond structures are general features of  hydration
of soluble proteins. Furthermore,
as noted in the statement above, the thermodynamic significance of such structures
hasn't been  established.

The XAFS results of Bowron, {\em et al.\/} \cite{BowronDT:Hydhfc}, which
are a motivation for this work, were directed toward an experimental
test of this clathrate concept.  Conditions were chosen to permit
formation of a Kr clathrate from the liquid solution in the sample cell.
Temperatures ranged from an  initial 5~C to -5~C with pressures of
approximately 110~bar.  Results taken before and after the transition
permitted comparison of the radial distribution of water oxygen atoms
conditional on a Kr atom. Those inferred radial distributions for the
liquid solution and the clathrate phase were {\em qualitatively\/}
different, as is discussed further below. This difference would be a
serious problem for a literal interpretation in terms of a clathrate
model of hydrophobicity.  A reasonable correspondence of radial solvent density certainly
should be a  fundamental requirement of a successful clathrate model and it
is a requirement that has received  less attention than orientational
distributions. 

Molecular-scale simulation of aqueous solutions of hydrophobic solutes
and of systems that form clathrates of common interest has been pursued
many times over several decades.  Much of that work on solutions is
referenced in \cite{PrattLR:Molthe}.  An early example of molecular
simulation of a methane clathrate is \cite{TSEJS:MOLSII}; a comparable
more recent example  can be found in \cite{ForrisdahlOK:MetchM} and  a
comprehensive review is given by Sloan \cite{Sloan}.  Conclusions from
that body of work are consistent with the results of the XAFS experiments
discussed above.  The radial layering of oxygen density conditional  on
the guest is much weaker in the  solution case.  Results from studies of
liquid solutions yield a modest maximum value of oxygen radial
distribution, typically slightly larger than 2, and with a weak minimum
that 
provides little physical distinction between first and second hydration
shells.  In contrast, the maximum values of g$_\mathrm{CO}$ in the
methane clathrate calculations are greater than 4 with an unambiguous
physical definition of the first hydration shell.   The important
work of Owicki \& Scheraga \cite{OWICKIJC:MONCIE} compared the radial
distributions of oxygen atoms neighboring a methane solute in liquid solution
to an ideal clathrate possibility.  In that case the mean inner shell
occupancy was about 23 and the correspondence with a clathrate case
seemed closest for the 24 vertex cage \cite{Sloan}, which larger than the 
smaller  20 vertex possibility presented by known clathrate crystals.  The radial
distribution was noted to be relatively diffuse in the solution case. 

Distributions of
orientational angles are often presented in solution  simulations and
sometimes compared explicitly to ideal results for clathrate crystal
geometries \cite{ALAGONAG:STRDAA}.  This question of the connection
between angle distributions and solution phase hydrophobicities was
addressed by an information theory analysis that arranged to include
systematically 2, 3, \ldots, n-body solvent correlations in
thermodynamic predictions of solution phase hydrophobic  hydration free
energies. It was a surprise that inclusion of just n=2 and 3-body
correlations {\em worsened\/} the thermodynamic predictions obtained
when only n=2-body correlations were used
\cite{PrattLR:Molthe,Gomez:99}.  That work suggested that an exclusive
focus on distributions of angles, which are obtained from 3-point
distributions, can be problematic for drawing conclusions about clathrate
models of hydrophobicity.  

It is helpful to keep  in mind the  observation discussed by Friedman and
Krishnan \cite{Friedman:73}: The sum of the solvation entropies of K$^+$ and
Cl$^-$ in water is roughly the same as twice the solvation entropy of Ar(aq), though
the case of methanol as solvent was different.  These hydration entropies are negative so solvent
ordering is  suggested.   But the specific molecular ordering of near neighbor water molecules
is expected to be qualitatively different in each of these cases.  So there is no simple,
unique inference of molecular structuring
available from the measured hydration entropy; furthermore, 
the observed entropies generally don't  imply that H-bonding
interactions among water molecules aren't substantially influenced by the  solute.

Still, clathrate pictorial interpretations are widely assumed.  Sloan
\cite{Sloan} notes that ``\ldots the water scientific community refers
to  short-lived cavities (not unit crystals sI, sII, or sH) as
clathrate-like structures in water.''  Indeed, it  is common to describe
the liquid solution results as `clathrate-like' without 
scrutiny of likeness or unlikeness to specific clathrate possibilities.

\begin{figure}[h]
\begin{center}
\includegraphics[width=5.0in]{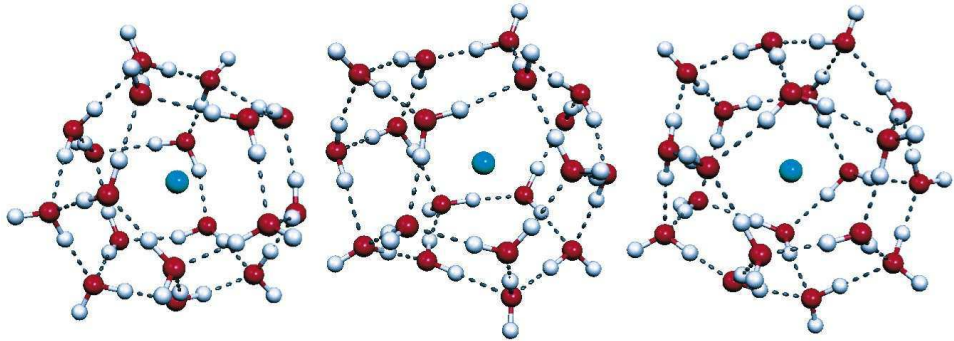}
\end{center}
\caption{Structures for Kr[H$_2$O]$_n$. H-bonds between the water molecules
are depicted by broken lines.  From left to right, the structures have
18, 19, and 20 water molecules, respectively.  Oxygen atoms are red, hydrogen atoms
are blue, and the krypton atom  is blue.}\label{fg:kr}
\end{figure}

\section{Perspective from {\em Quasi-Chemical\/} Theory}

Quasi-chemical theories are built upon a distinction between defined
inner  and outer shell regions \cite{Paulaitis:APC:02,Pratt:ES:99}.  The
intention is to define an  inner shell that  can be studied at molecular
resolution using tools of modern computational chemistry. For the outer shell region, in contrast, 
genuine statistical mechanics can't be avoided but it ought to be simpler because
complicated chemical interactions aren't a direct  concern there. The inner shell
is defined geometrically \cite{HummerG:Newphe} by specification of an
indicator function $b(\mathbf{j})$  that is one (1) when solution
molecule $\mathbf{j}$ occupies the intended inner shell and zero (0)
otherwise.   Fig.~\ref{fg:kr} shows candidates for inner shell complexes
Kr(H$_2$O)$_n$ and suggests how this approach might apply to build
statistical thermodynamic models starting from `clathrate-like'
concepts.

With the inner shell defined, the interaction contribution to the
chemical potential of the Kr solute can be expressed as
\cite{HummerG:Newphe,Paulaitis:APC:02}
\begin{eqnarray}
\beta \mu_{\mathrm{Kr}(aq)}^{ex} =& - &\ln \left ( 1+\sum_{m\ge1} K_m
 \rho_{\mathrm{H}_2{\mathrm{O}}}{}^m  \right)  \nonumber \\
 & - &\ln \left< \left< e^{-\beta \Delta U_{\mathrm{Kr}}} \Pi_j
 \left(1-b({\mathbf{j}})\right) \right> \right>_0~.
\label{inner_outerK}
\end{eqnarray} 
($\beta^{-1}$ = kT, with T the temperature and k the Boltzmann
constant.) This excess chemical potential is partitioned into an inner
shell contribution (the first term on the right of
Eq.~\ref{inner_outerK}) and an outer shell contribution (the remainder).
 The inner shell contribution is associated with equilibrium of chemical
equations
\begin{equation} 
\mathrm{Kr} + n \mathrm{H}_2\mathrm{O}  \rightleftharpoons
{\mathrm{Kr}(\mathrm{H}_2\mathrm{O})_n}~
\label{equilibrium}
\end{equation}
for formation of inner shell complexes, as in Fig.~\ref {fg:kr}, {\it in
situ.\/}  The coefficients K$_m$ are equilibrium ratios for these
equations.

The coefficients K$_m$  might be obtained by observations of the
population of Kr atoms coordinated to $m$ water molecules, $m$=0,1,\ldots
This directs attention to  the probability that a distinguished Kr atom
has $m$  inner shell water molecule ligands. Those fractions will be denoted
by $x_m$ and  are given by
\begin{eqnarray}
x_m = {K_m \rho_{\mathrm{H}_2{\mathrm{O}}}{}^m \over 1+\sum_{n\ge1} K_n
\rho_{\mathrm{H}_2{\mathrm{O}}}{}^n}~.
\label{xm}
\end{eqnarray}
The probabilities $x_m$ are physical observables that could be obtained,
for example, from a
simulation. Our first goal is to  investigate these $x_m$ to see whether
they give any suggestion of coordination species present in
corresponding known clathrate hydrates.   We note these distributions,
$x_m$, were conventionally considered in historical computer simulations
for these problems;  they were then referred to as `quasi-component'
distributions \cite{SWAMINATHANS:MONSSD}.  This  discussion attempts to
situate these distributions in helpful statistical thermodynamic
theories.

The natural initial implementation of these quasi-chemical ideas,
particularly to ion hydration, has been to treat effects of material
external to the cluster by an approximation 
\begin{eqnarray}
K_m \approx
\exp\left[{-\beta \left(
\mu_{\mathrm{Kr}(\mathrm{H}_2\mathrm{O})_m}^{ex}-m
\mu_{\mathrm{H}_2\mathrm{O} }^{ex}\right)}\right] K_m{}^{(0)}
\label{pqca}
\end{eqnarray}
 where
$K_m{}^{(0)}$ is the  equilibrium ratio for Eq.~\ref{equilibrium} in a
dilute gas.  The required hydration free energies are obtained
standardly from a dielectric continuum model \cite{grabowski} and includes a contribution from the
outer shell term corresponding to the bare ion.  This approach will be used in the
results below.
and the required hydration free energies are obtained standardly from
a dielectric continuum model (37) and includes a contribution from the
outer shell term corresponding to the bare ion.  This approach

Recently it has been indicated in Ref.~\cite{Paulaitis:APC:02} how these inner and
outer shell contributions can be naturally recombined after the most
natural primitive approximations. In  that sense, it isn't necessary to
consider these two terms separately. That development leads further to a
revised approximation scheme in which $K_m \approx \gamma^m K_m{}^{(0)}$,
with $\gamma$ a Lagrange multiplier that plays the same role here as an
activity coefficient.  Further, -kT$\ln\gamma$ can also be seen as an
effective external  field operating on each ligand molecule, analogous
to the prefactor of $K_m{}^{(0)}$ in Eq.~\ref{pqca}.  This leads finally
to the concept of a self-consistent field treatment of quasi-chemical
approximations \cite{ashbaugh:02}.  

 \begin{figure}[h]
 \begin{center}
 \includegraphics[width=5.0in]{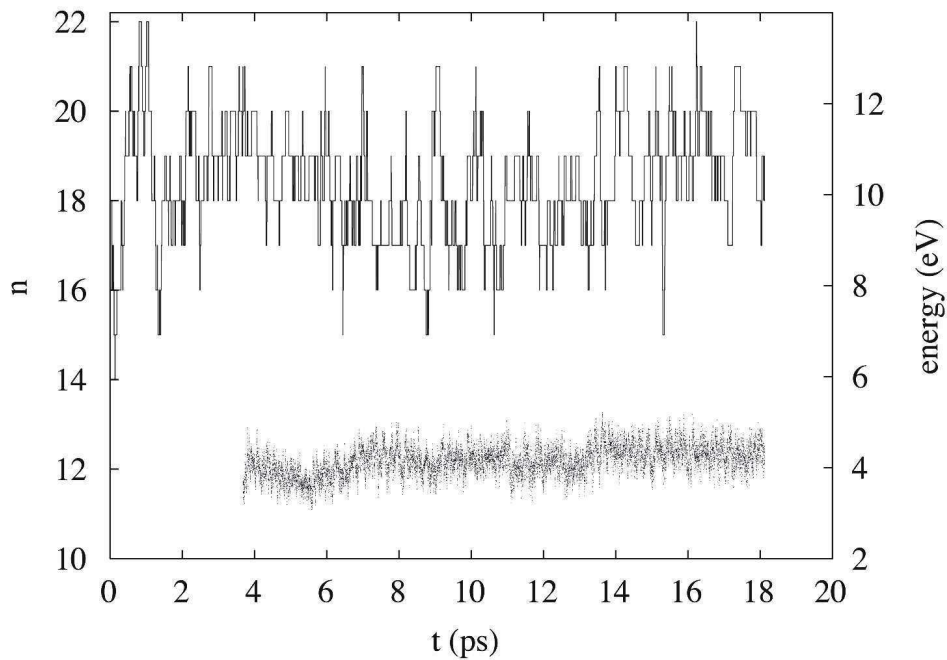}
 \end{center}
 \caption{Oxygen coordination of the Kr atom as a function of time
 (upper trace and left axis).  The  total kinetic energy of the atomic
 motion shown
in  the lower trace demonstrates the stability of the
 temperature state.}
 \label{fig:bonds}
 \end{figure}

\section{{\em Ab Initio\/} Molecular Dynamics for Kr(aq) and 
Comparison with XAFS Results}

The system simulated by AIMD consisted of one Kr atom surrounded by 32
water molecules in a cubic box with sides of length 9.87 {\AA} and
periodic boundary conditions.  An initial structure was obtained
utilizing results from an earlier study of Li$^+$(aq)
\cite{RempeSB:hydnL+,RempeSB:hydnN+}.  A Kr atom was substituted for the
Li$^+$ ion in a configuration from that calculation, the full system was
then optimized in a much larger cell and replaced in the original
simulation cube.  Optimization of configurations drawn from the 
subsequent molecular dynamics trajectory and variations of total
optimized energies with the simulation cell size suggests that this
system is at a realistically low pressure.  A more precise
estimate of the pressure was not possible in the present effort.

\begin{figure}[h]
 \begin{center}
 \includegraphics[width=5.0in]{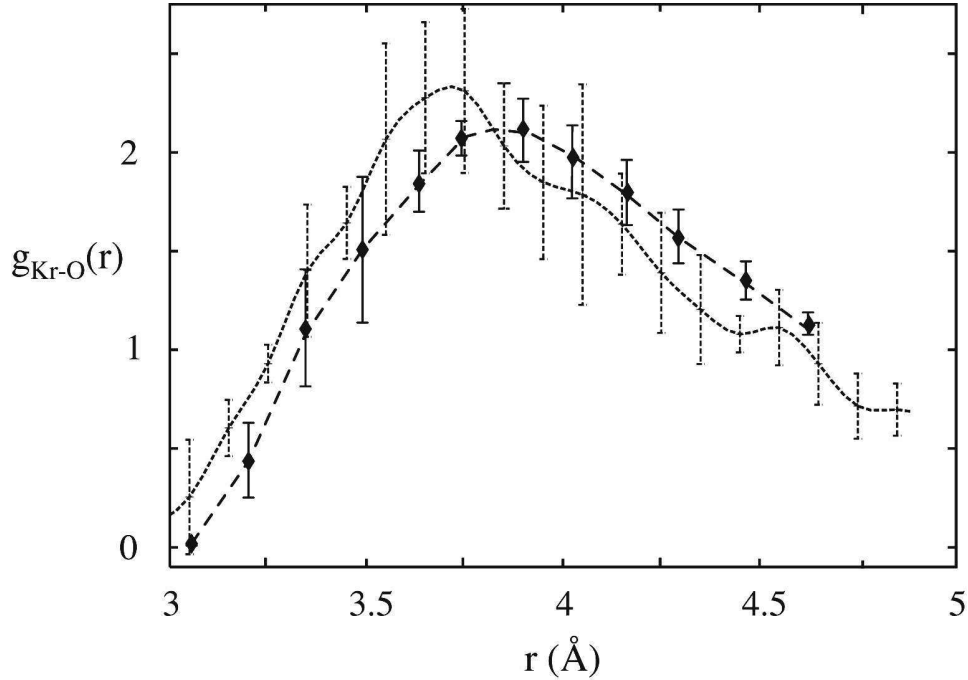}
 \end{center}
 \caption{Pair distribution function for Kr-O measured by experiment
 (dashed line) and by AIMD simulation (solid line).  In the latter case,
 indicated error bars of $\pm 2 \sigma$ were estimated on the
 assumption that the last four 2.5 ps trajectory segments are
 independent.  Error bars of $\pm \sigma$ are plotted with the
 experimental data.  }
 \label{fig:gKrOcomp}
 \end{figure}

The calculations based upon an electron density functional
description\cite{perdew:91} of the electronic structure and interatomic
forces were carried out on the Kr(aq) system utilizing the VASP
program\cite{vasp:93, vasp2:96}.  The ions were represented by ultrasoft
pseudopotentials\cite{vanderbilt, vasp:94}, in the local density
approximation for Kr and in the gradient-corrected approximation for O
and H, and a kinetic energy cutoff of 29.10 Ry defined the plane wave
basis expansions of the valence electronic wave functions.  The valence
electrons consisted of eight (8) electrons  for Kr (4s and 4p), six (6)
electrons for O (2s and 2p), and one (1) electron for H (1s). The
equations of motion were integrated in time steps of 1 fs with a
thermostat set at 300 K during the first 4 ps of simulation time. The
thermostat was then abandoned; the temperature rose and settled at
341$\pm$24~K (see  Fig.~\ref{fig:bonds}), higher than the experiments of
interest here. The precision of temperature characterization and the
corresponding awkwardness in temperature adjustment is a practical
limitation of these simulations that treat small systems over short
times.  Subsequent work should investigate temperature effects on these
simulations, but we will proceed here in considering the results.

The structural analyses presented here apply to the last 10 ps of the 18
ps trajectory.  During this  time, the initial structure with hydration
number $n=19$ relaxed into structures with $n=15$ to $n=22$ nearest
neighbors within a radius of 5.1~\AA; see Fig.~\ref{fig:bonds}. The
resulting Kr-O radial distribution function, plotted with the solid line
in Fig.~\ref{fig:gKrOcomp}, shows a build-up of water density to
slightly more than double the bulk value, g$_{max}$(r)$\approx$2.4 at a
radius r$\approx$3.7{\AA} from the Kr atom. A comparison with the
experimental XAFS data produced by \cite{finney_prl:97}
(Fig.~\ref{fig:gKrOcomp}) shows an overlap of error bars between the
calculated and the experimentally determined radial distribution
functions.   The maximum value, g$_{max}$(r)$\approx$2.4, is about the
same as the value  found from a force field model adjusted to give
experimental solution thermodynamics
\cite{SWOPEWC:MOLMCS,WATANABEK:MOLSTH} and is slightly larger than the
result found by \cite{AshbaughHS:RSPM} for the case of a corresponding
sized hard sphere solute in SPC water, which is, as shown there, also in
good agreement with the revised scaled particle model
\cite{Stillinger:73}.

\begin{figure}
 \begin{center}
 \includegraphics[width=4.0in]{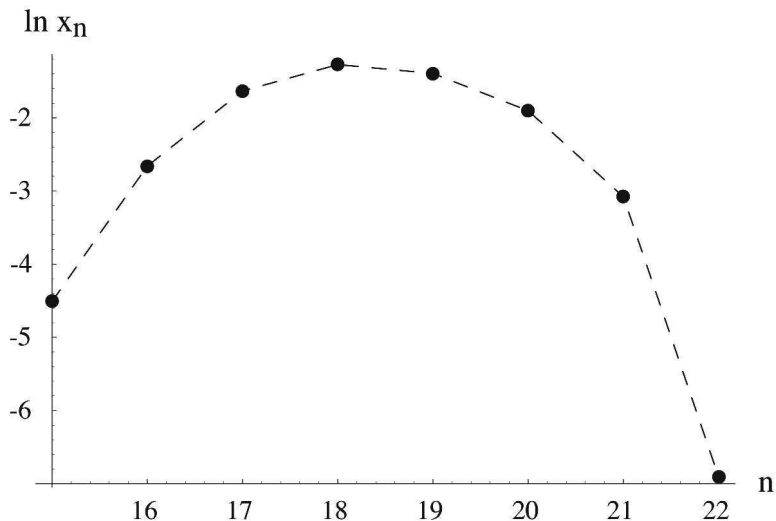}
 \end{center}
 \caption{Probability distributions of inner-shell structures with $n$
 water molecules surrounding Kr atom where the inner-shell boundary is
 defined at r=5.10~{\AA}. }
\label{fig:xn}
 \end{figure}

The average occupancy of the  inner shell defined by the radius
r$\le$5.1\AA\ is $\langle n \rangle$ =18$\pm$1 and n=18 is found also to
be the most probable structure, as is shown  in Fig.~\ref{fig:xn}. The
n=20 case  is about half as likely as the  most probable n=18 case and
either n=19 or n=17 is more probable  than  n=20 by these results. This
$x_n$  distribution of coordination numbers is unimodal. In a limited
range of thermodynamic states approaching a clathrate phase boundary,
the $x_n$ distribution  might be expected to exhibit magic number
features reflecting metastable coordination possibilities
\cite{LONGJP:QUAWCA}.  Those possibilities are not observed here.

\begin{figure}[h]
 \begin{center}
 \includegraphics[width=5.0in]{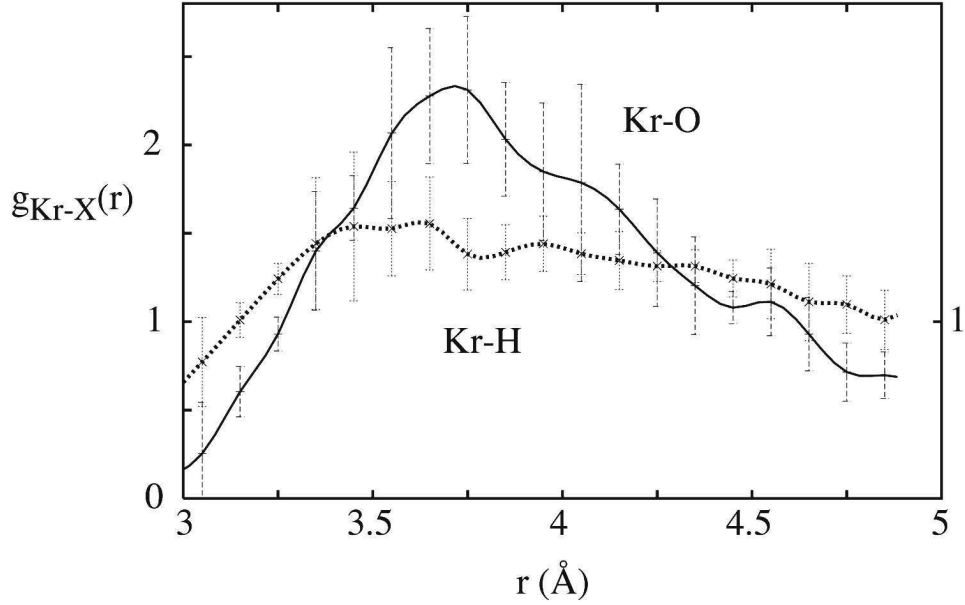}
 \end{center}
 \caption{Pair distribution function for Kr-O and Kr-H determined
by AIMD simulation.  Error bars indicate the estimated 95\% confidence level.  }
 \label{fig:gKrOH}
 \end{figure}

Consideration of the radial  Kr-H  pair distribution function
(Fig.~\ref{fig:gKrOH}) shows a featureless  distribution.  These results
are not dissimilar to recent computations of radial hydrogen
distributions conditional on a hard sphere solute in SPC water
\cite{AshbaughHS:RSPM}; although those calculations have better spatial
resolution and thus additional observable structure, the maximum
hydrogen density is about the same here.

Although these calculations have clear limitations of scale and thus
correspondence to the experiment, to within the errors of the
theoretical and experimental measurements,  {\em they do corroborate the
experimentally observed local fluid structure.}  They provide also
the more probable features  of the coordination number distribution,
$x_n$,  of physical interest from the point of view of quasi-chemical
models.  The  coordination structures observed in the AIMD simulation
should provide an appropriate starting point in considering  inner
shell coordination species that might found quasi-chemical models.

The following sections  consider the  outer   and inner shell
contributions  to the hydration free energy of Kr in turn.

\section{Outer Shell Contributions}
		
The outer shell term would be the hydration free energy under the
hypothetical constraint  that inner shell occupancy be prohibited. We
anticipate that an electrostatic contribution, estimated with a
dielectric continuum model, will be included in the  modified equilibrium
ratios obtained below; see Eq.~\ref{pqca} and the supporting discussion.
For the  Kr case considered here,  that leaves packing interactions
associated with hard spherical exclusion volume and
weaker London dispersion interactions.  We will discuss those outer shell
contributions here.

Scaled particle theory provides a connection between outer shell
contributions to the free energy and the packing of water about an
evacuated cavity.  The free energy for emptying the
inner sphere is given by the work of growing a hard sphere solute in
aqueous solution
\begin{eqnarray}
\beta \mu ^{ex}(\mathrm{outer\ shell})=4\pi \rho_{\mathrm{H}_2{\mathrm{O}}} \int\limits_0^R {G(\lambda
)\lambda ^2d\lambda }
\label{spt}
\end{eqnarray}
where $R$ is the inner shell radius,  and $G(\lambda)$ is the contact
value of the cavity-water oxygen radial distribution function as a
function of the distance of closest approach $\lambda$.  This expression
can be thought of as the $p\cdot dV$, $\rho_{\mathrm{H}_2{\mathrm{O}}} 
k T  G(\lambda) \cdot 4\pi\lambda^2d\lambda$, work associated with
growing an empty cavity into solution.  The surface tension associated
with differentially increasing the surface area of the outer shell
cavity is given by the derivative of Eq.~\ref{spt} with respect to outer shell
surface area
\begin{eqnarray}
\beta\gamma (R) =\left( {{{\partial \beta\mu } \over {\partial R}}} \right){{\partial 
R} \over {\partial \left( {4\pi R^2}
\right)}}=\frac{1}{2}\rho_{\mathrm{H}_2{\mathrm{O}}}G(R)R
\label{g1}
\end{eqnarray}
This derivative has also been approximated simply by the ratio of the
hydration free energy of exclusion volume to its surface area
\begin{eqnarray}
\beta\gamma (R) \approx  \rho_{\mathrm{H}_2{\mathrm{O}}}\int\limits_0^R {G(\lambda
)({\lambda\over R}) ^2d\lambda }
\label{g2}
\end{eqnarray}
These two distinct expressions, Eq.~\ref{g1} and  \ref{g2}, yield
different values for the surface tension for submacroscopic cavities.

The contact function $G(\lambda)$ is readily obtained from molecular
simulations of hard sphere hydration  at discrete values of $\lambda$.
Alternatively, Stillinger proposed \cite{Stillinger:73} a functional form for $G(\lambda)$
that interpolates between the known microscopic and macroscopic limiting forms:
\begin{eqnarray}
G(\lambda )=\left\{ {\matrix{{{{1+\left( {\pi \rho /\lambda }
\right)\int\limits_0^{2\lambda } {g_{OO}(z)(z-2\lambda )z^2dz}} \over
{1-4\pi \rho \lambda ^3/3+(\pi \rho )^2\int\limits_0^{2\lambda }
{g_{OO}(z)(z^3/6-2\lambda ^2z+8\lambda ^3/3)z^2dz}}},\ \ \lambda <1.75\mathrm{\AA}}\cr
{p_{sat}/\rho kT+{{(2\gamma _{lv}/\rho kT)} \over \lambda }+{{G_2} \over
{\lambda ^2}}+{{G_4} \over {\lambda ^4}},\ \ \lambda >1.75\mathrm{\AA}}\cr }} \right.
\label{rspm}
\end{eqnarray}
where $g_{OO}(r)$ is the water oxygen-oxygen radial distribution
function, $p_{sat}$ is the saturation pressure of water in equilibrium
with its vapor, $\gamma_{lv}$ is the vapor-liquid surface tension.  The
coefficients $G_2$ and $G_4$ are chosen so that the contact function is
seamless, matching $G(\lambda)$ and its first derivative at
$\lambda$=1.75\AA, different from the originally suggested 1.95\AA\
\cite{Stillinger:73,AshbaughHS:RSPM}. Extensive molecular simulations
\cite{AshbaughHS:RSPM} have verified Stillinger's functional form for
water and give confidence for calculating outer shell contributions to
the hydration of Kr.  

\begin{figure}[h]
\begin{center}
\includegraphics[width=5.0in]{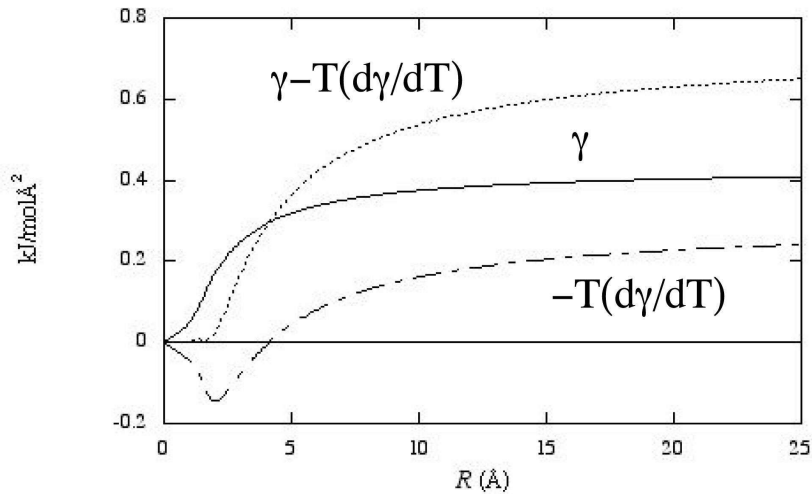}
\end{center}
\caption{Apparent surface tension $\gamma$, and its temperature
dependence, predicted by the revised scaled particle model Eq.~\ref{g1}
at 25~C, as a function of inner shell cavity radius. Where the solid
curve exceeds the dotted curve, entropic contributions increase the
value of this apparent surface tension.}\label{apparent_gamma}
\label{gammaT}
\end{figure}

To this end we have used simulation results for
the radial distribution function of SPC/E water along the saturation
line (Prof. S. Garde - personal communication) with the experimental
surface tension of water and the corresponding surface hydration
properties to evaluate $G(\lambda)$ using Eq.~\ref{rspm}. The resulting surface
tension as a function of  cavity size at 25~C is shown in
Fig.~\ref{gammaT}. While Eqs.~\ref{g1} and \ref{g2} for the surface
tension are zero for a cavity of zero size and converge to the same
ultimate value for an infinite cavity ($\gamma$ = 0.411 kJ/mol-\AA$^2$ =
72 dyn/cm) assuming the saturation pressure of water is negligible, they
differ at intermediate molecular sizes.  In particular, Eq.~\ref{g1}
nearly attains its plateau value for cavities on the order of 5 to
10~\AA\  in radius, while Eq.~\ref{g2}  more slowly approaches the large
sphere limit and is still less than 0.4 kJ/mol-\AA$^2$  for the largest
sizes considered.  Particularly interesting is the temperature
dependence of the outer shell cavity hydration free energy;
Fig.~\ref{gammaT} shows a distinct negative entropy for hydrophobic
hydration for molecularly sized cavities that dominates the hydration
free energy and is consistent with the known thermodynamic
characteristics of small hydrophobe hydration.  With increasing cavity
size the entropy changes sign and becomes favorable with increasing
size, approaching the surface free energy of a flat surface.  Drying
phenomena are, of course,  not a separate issue for this revised scaled
particle model\cite{PrattLR:Molthe,Lum:JPCB:99,PrattLR:Quatst}.  

The formula Eq.~\ref{spt} gives a  free energy contribution of 9~kcal/mol which represents the work of rearranging water to grow an empty cavity of radius R=3.9\AA\  corresponding approximately to the distance of the inner shell water molecules in minimum energy cage structures discussed below. The first-order perturbation theory estimate
\cite{allen-tildesley} of the effects of dispersion interactions between
the Kr  solute and outer shell water molecules is
-$\frac{16\pi\epsilon}{3}(\frac{\sigma}{R_0})^3\rho_W\sigma^3$; with
$\epsilon$= 0.19~kcal/mol,  and $\sigma$=3.38\AA, the Lennard-Jones
parameters for Kr-W van~der~Waals interactions, this is roughly
-3~kcal/mol with $R_0$=3.9\AA.

We conclude this  discussion by emphasizing the important point that
this treatment of the outer shell  term gives a positive contribution,
roughly 6~kcal/mol.  The experimental  value for the
hydration free energy of Kr(aq) is roughly
2~kcal/mol \cite{kr-data}. Although  these observations are
straightforward, the non-trivial implication  here is that inner shell
contribution must be {\em negative\/} and of  comparable size.

\section{Inner Shell Structures and Contributions}

The advantage of the quasi-chemical formulation is that  Eq.~\ref{pqca}
is a reasonable initial approximation to the equilibrium constants for
complex formation in solution, the quantities sought for the inner shell
contribution to the solute hydration free energy. The harmonic
approximation to $K_m{}^{(0)}$ is available from electronic structure
calculations.
	
Fig.~\ref{fg:kr} shows inner shell structures that have been obtained
and we note some basic observations. Firstly, these complexes are
large enough that satisfactory electronic structure calculations are
non-trivial\cite{LudwigR:Quacet}.  Secondly,  finding mechanically
stable structures for these complexes is non-trivial, particularly for
values of n$<$18. For example, extracting inner shell structures  from
the AIMD trajectory and then energy  optimizing always resulted in
decomposition, {\em  i.e.,\/} some of the water molecules were
repositioned to the outer shell and then $b(\mathbf{j})$=0 for some
$\mathbf{j}$.  Finding mechanically stable complexes required  some
subtlety, as described below, and still only n$\ge$ 18 complexes were obtained.

A 20-vertex pentagonal dodecahedron cage, present as the small cavity
in all known clathrate hydrate structures \cite{Sloan},  was built enclosing a
Kr atom in several steps. First we built a carbon cage and minimized
that structure with molecular mechanics. Then each vertex of this
dodecahedron was replaced by oxygen atoms and hydrogen atoms were added
to give a cage composed of water molecules with the Kr embedded inside
it. This structure was energy minimized with a molecular
mechanics potential for Kr-water interactions. The minimization occurred
over multiple stages in which the force constant of a harmonic force
restraining the oxygen atoms was progressively decreased. The final
structure obtained with a low force constant (5~kcal/mole/{\AA}$^2$) was
then subjected to quantum chemical optimization. The n=19 and n=18
structures were obtained by removal of water molecules from the
optimized n=20 structure. For the n=18 structure, we
had to once again go through the routine of employing molecular
mechanics before any cage-like structure could be obtained.

Limited computational resources permitted only optimizations using 
HF/6-31G theory. The quantum chemical optimizations first employed
steepest descent to  relax the molecular mechanics structure rapidly 
with the final geometry obtained using the Newton-Raphson procedure. All
calculations were performed using Gaussian  \cite{gaussian}. Frequencies
(and zero-point and thermal contributions to the free energy) were
computed at the HF/6-31G level of theory. Non-negative curvatures
confirmed a true minimum. These structures (Fig.~\ref{fg:kr}) are
clearly metastable  and the difficulty in finding these minima suggests
that the catchment volumes are small.

\begin{table}[h]
\caption{$\Delta E$, electronic energy change upon complexation. $\Delta G_{corr}$, zero-point and thermal corrections upon complexation. $\Delta G_{298}$ is the free energy change accounting for the actual density of water, that is the free energy change at 1~atm. pressure is adjusted by -nRT ln (1354). $\Delta \mu$ is the change in the solvation free energy. $\Delta G$(aq) is the excess chemical potential of Kr for the assumed hydration structures. All values are in kcal/mole. }\label{summary} 
\begin{center}
\begin{tabular}{crrrrr}\hline
   n  & $\Delta E$ & $\Delta G_{corr}$ & $\Delta G_{298}$ & $\Delta \mu$ & $\Delta G$(aq) \\ \hline
 18  &  -195.6        &    220.6                   &         -52.4               &        93.0       &     40.6   \\
  19 &  -204.2        &    226.6                   &          -59.3              &        98.4       &     39.1   \\
 20 &  -218.9        &    245.9                    &         -59.0               &       105.7      &     46.7 \\ \hline
\end{tabular}
\end{center}
\label{tb:summary}
\end{table}

For the single point energies we used the B3LYP hybrid density
functional, but with two different basis sets. This functional doesn't
account for  Kr-H$_2$O interactions of London dispersion character; but
it does treat electrostatic  induction phenomena and is adequate for the
water-water interactions. We will address this omission of Kr-H$_2$O
dispersion interactions later.  The basis sets were 6-31+G(d,p) for all
atoms or the 6-31G for Kr and a much larger 6-311+G(2d,p) basis for the
O and H atoms. Only the results with the latter choice are presented
here (see Table~\ref{tb:summary}).

The electrostatic contributions required by Eq.~\ref{pqca} for the
Kr[H$_2$O]$_n$ cluster and the individual water molecules were computed
using a continuum dielectric model. The exterior of the system was
assigned the dielectric constant of liquid water and the interior had a
dielectric constant of 1. The calculations were performed using a
boundary integral formulation of the governing equations
\cite{lenhoff:jcc90}, with the surface tessellation obtained using the
MSMS program \cite{sanner}. The partial atomic charges needed in this
calculation were ESP charges obtained using the ChelpG procedure in
Gaussian \cite{gaussian}. For the charge fitting procedure, the Gaussian
default radii for O (1.75{\AA}) and H (1.45{\AA}) atoms were used. For
Kr a radius of 2.0{\AA} was used. A radius of 3.0{\AA} for Kr leads to
an unphysically high positive charge on the Kr atom and, therefore, was
not used. The solvation free energies are, in fact, insensitive to this
choice and the values differ by only about 0.6~kcal/mole. 
Table~\ref{tb:summary} summarizes the inner shell results.

In the calculations above, the  zero-point energies were scaled by a
factor of 0.9135. This factor is suitable for the HF/6-31G(d) level and
has been used  for the present HF/6-31G level calculations. For the
cases that  we could build closed cages, we included an approximate
correction for residual entropy corresponding to conformational
multiplicity. Assuming ideal tetrahedral coordination for each water,
this residual entropy amounts to $S/k\approx\ln(3/2)^n$
\cite{pauling:re}, where $n$ is the number of water molecules in the
cage.  Further, MP2 level calculations that do treat dispersion interactions between
an isolated Kr-water pair
suggest a correction that amounts to about
-0.3~kcal/mole for each Kr-water pair; the excess chemical potentials
would need to be adjusted by this amount also. Thus for the n=18 case,
the excess free energy of Kr is 30.9~kcal/mole and 35.9~kcal/mole for
the n=20 case. With or without these additional corrections, it is clear
that the inner shell contribution is in the range of 30-50~kcal/mole,
which is an order of magnitude greater than the experimental value of
around 2~kcal/mole. Note that accuracies of 2~kcal/mole are barely
achievable with all but the highest level of electronic structure
methods.

Considering the balance  of inner and outer shell contributions, we
return to the fact that the accurately known  outer shell contributions
discussed above are positive and larger than the thermodynamical value.
This  implies that the inner shell contributions must be negative and
about as large  in magnitude as the experimental  value. Here our 
estimates for that inner shell contributions are positive and large;
this qualitative situation is unlikely to change substantially with more
sophisticated electronic structure calculations. We conclude that the
effects of the medium external to our clusters are decisive in
stabilizing these structures.   They must help  to pull open and
stabilize the cage. 
This physical realization is consistent with
Sloan's view that these structures should be visualized as  {\em dandelions\/}
rather than spherical complexes \cite{Sloan}.   

More specifically, the results of Table~\ref{tb:summary}  
suggest that the the dielectric model used here to describe
the influence of the outer sphere material, though satisfactory
for some corresponding ion  hydration problems, is not satisfactory
for the present application.  It is possible that the  effective fields, hinted following
Eq.~\ref{pqca}, which are self-consistent with the  known fluid
densities and reflect the influence of outer shell material on the
$K_n$'s \cite{ashbaugh:02,PrattLR:Quatst}, should describe these effects
more satisfactorily, but resolution of that possibility will remain for
subsequent work.

\section{``What are we to tell students?''}

Work  on the problem of  hydrophobic effects has the maturity of many
decades of effort; that time scale certainly reflects the difficulty of
this theoretical problem. A conceptualization reflecting that maturity
might be subtle, expressing the complexity of water as a
material, but can be simple.   An example of a surprising  simple result is the
formula
\begin{eqnarray}
\mu_{\mathrm{Kr}(aq)}^{ex}=-A\rho_W(T) + BT\rho_W(T)^2 +C T~,
\end{eqnarray}
with parameters $A$, $B$, and $C$, and $\rho_W(T)$ is the density of liquid water coexistence with
its vapor at temperature $T$ \cite{PrattLR:Molthe}.  That this formula describes signature hydrophobic thermodynamic properties for small hydrophobic solutes 
was unanticipated  \cite{Garde:PRL:96}, but it would have been unreasonable
to hope for a simpler  result.  This simple formula  points to  the distinctive
equation of state of liquid water as a primary  feature of hydrophobicity.

A recent attempt at an answer to ``what are we to tell
students'' appeared in \cite{PrattLR:CR02}.  That attempt tried to
produce a faithful  physical paraphrase of the recent molecular theories
which, though non-committal on simple structural hypotheses, have
identified a simple explanation of entropy convergence of
hydrophobicities \cite{Garde:PRL:96,GardeS:Temdhh,AshbaughHS:simmtt}.
Here we refine and thus  simplify that previous answer.

Several points can be made in  preparation for this discussion. Firstly,
we note that the revised scaled particle model \cite{Stillinger:73}  is
the most successful theory of hydrophobicity for spherical solutes
\cite{AshbaughHS:RSPM,Pratt:PNAS:92,Palma}. The widely recognized point
that ``\ldots the orientational arrangement of vicinal water molecules,
is absent from the theory'' \cite{Guillot:JCP:93} does {\em not\/} mean
that these theoretical approaches are specifically approximate because
of a neglect  of orientational effects.  Though the truth-content of
this assertion has never changed, reappreciation of molecular theories has
made this point explicit over recent years
\cite{PrattLR:Molthe,HummerG:Newphe,Pratt:NATO99}. In fact, the revised
scaled particle model  along with the Pratt-Chandler theory
\cite{Pratt:JCP:77}, the information theory  models
\cite{HummerG:Newphe,Gomez:99}, and quasi-chemical treatments
\cite{PrattLR:Molthe} specifically do not neglect orientational effects.
What  is more, because these theoretical models firmly incorporate
experimental information \cite{Pratt:NATO99}, they do  not assume that
H-bonding among water molecules is non-cooperative, as is the case for 
most computer simulations of aqueous solutions.

Secondly, hydrophobicity as judged by hydration free energy is greatest
at moderately elevated temperatures $>$100~C along the vapor
saturation curve,  as was emphasized by \cite{Murphy:90}.  This
observation may be non-canonical
\cite{Dill:90a,Murphy:90:r,Herzfeld:91}; but the unfavorable hydration
free energy of simple hydrophobic species is an objective property of 
those systems and is largest in these higher temperature thermodynamic
states.  What is more, the most provoking puzzle for  molecular
mechanisms of hydrophobic phenomena has always been the apparent
increase in attractive strength of hydrophobic effects with increasing
temperature for temperatures not too high.  This point is experimentally
clear in the  phenomena of cold-denaturation wherein unfolded soluble
proteins fold  {\em upon heating}.  [See, for example, \cite{Li:2001.b}.
Proteins {\em are\/} complex molecules, but in that elastin example no
hydrophilic side chains complicated the considerations in that way.  And the
experimental temperature for collapse upon heating --- 27~C ---
is in the stable range of  the liquid phase of water.]  This point  is
conceptually troublesome because the molecular structural pictures,
including clathrate models, seem to point to low temperature regimes and
behaviors as identifying the essence of hydrophobicity.  The discussion
of \cite{STILLINGERFH:WATR}, which suggested connections between the
behaviors of supercooled liquid water and hydrophobic effects, provides
an example of those intuitions.

There are two characteristics of liquid water that are featured in the
new answer that we propose.  The first characteristic is that the liquid
water matrix  is stiffer  on a molecular scale than are comparative
organic solvents when  confronted with the excluded volume of
hydrophobic molecular solutes
\cite{PrattLR:CR02,HummerG:Newphe,Palma,Hummer:JPCB:98}.  An indication
of this relative stiffness is provided by comparison of  experimental
compressibilities of the usual  solvents \cite{PrattLR:Molthe}.  We
don't give a structural picture for that relative stiffness, but it is
due to  intermolecular interactions among solvent molecules, H-bonding
in the  case  of liquid water \cite{HEADGORDONT:ORIPPL}.  This stiffness
is the principal determinant of the low solubility of inert gases in
liquid water.  Furthermore, this stiffness is  weakly temperature
dependent in the case of liquid water. This temperature insensitivity is
again suggested  by consideration of the isothermal compressibility of
water; the temperature variation of that isothermal compressibility
displays a minimum at 46~C and low pressure.  In theories such as
Eq.~\ref{rspm}, this stiffness and its temperature dependence is carried
by the experimental $g_{OO}(r)$, which is distinctive of liquid water.

[We note in passing, that the liquid-vapor interfacial tension and its
temperature dependence, $\gamma_{lv}(T)$, appears also in the model
Eq.~\ref{rspm}.  In previous scaled particle models where this empirical
information was not specified, the temperature dependence of the implied
 $\gamma_{lv}(T)$ was found to be
improper\cite{Stillinger:73,Ben-Naim:67}, though the magnitude of this
parameter in the temperature region of most interest was reasonable.  In
some other  important aspects that more primitive scaled particle model
was satisfactory\cite{Pierotti:63,Pierotti:76}. Thus, a reasonable view
is that the specific empirical temperature dependence of
$\gamma_{lv}(T)$ better resolves the conflicting temperature
dependences, but is secondary to the classic hydrophobic temperature
dependences that are clearest for submacromolecular scales (see
Fig.~\ref{gammaT}).]

The second characteristic in our  answer is the variation of the liquid
density along the liquid-vapor coexistence curve in the temperature
regimes of interest here \cite{AshbaughHS:simmtt}.  The critical
temperature of liquid water is significantly higher than is the case for
the comparative organic solvents. The coefficient of thermal expansion
along the coexistence curve, $\alpha_\sigma$
\cite{Rowlinson:Swinton:82}, is typically more than five times smaller
for water than for common organic solvents. It  is a secondary curiosity
that liquid water has a small regime of density increase with increasing
temperature; we are interested here in a much broader temperature
region.  Nevertheless, the densities of typical organic solvents
decrease more steadfastly with increasing temperature than does the
density of water.

These two points lead to a picture in which the aqueous medium is
stiffer over a  substantial temperature range and expands with
temperature less significantly than the natural comparative solvents. If
these structural features of the aqueous medium are thus buffered
against normal changes  with increasing temperatures, then at higher
temperatures the solvent exerts a higher kinetic pressure through
repulsive collisions with hydrophobic solutes which don't experience
other interactions of  comparable significance. These collisions  are
proportionally more energetic with increasing temperature and the
aqueous environment thus becomes more unfavorable for hydrohobic solutes
with increasing temperature. The rate of density decrease with
increasing temperature eventually does dominate this mechanism  at the
highest temperatures of interest here, $>$100 C, and less unconventional
behavior is then expected.  This is our response to  ``what are we to tell
students?''

If  a balancing act  minimizing the variations of the structure of the
aqueous medium with temperature is possible, it should be a useful trick
since it should have the consequence of expanding the temperature window
over which biomolecular structures are stable and functional
\cite{PrattLR:CR02}.

We emphasize again, as in Ref.  \cite{PrattLR:CR02}, that hydrogen bonding,
tetrahedrality of coordination, random networks and related concepts are
not direct features of this answer. Nevertheless, they are relevant to
understanding liquid water; they are elements in the bag of tricks that
is used to achieve the engineering consequences that are discussed in
the picture above.

\section{Conclusions}

The coordination number of Kr in liquid water solution is significantly
different from known clathrate hydrate phases.  These coordination
numbers play a direct role in quasi-chemical descriptions of the
hydration thermodynamics.  Thus, the coordination number differences
between liquid  solution and clathrate phases can be expected to lead 
to substantial thermodynamic effects. In these respects, a clathrate
picture is not supported by our current information on the molecular
scale hydration of Kr(aq); the euphemism `clathrate-like' has to be
understood as `clathrate-unlike' in these respects.

The present results for the distribution of Kr(aq) coordination numbers
give no suggestion of lower  probability, metastable, magic number
coordination structures
that might reflect the coordination possibilities in a known, lower temperature,
clathrate hydrate phase; see Fig.~\ref{fig:xn}.

The isolated inner shell Kr[H$_2$O]$_n$ complexes that have been
obtained for n=18, 19, and 20 have a delicate stability. When inner shell
coordination structures were extracted from the simulation of the liquid,
and then subjected to quantum chemical optimization, they decomposed.
 Inner shell
complexes for n$<$18 were not  found here and therefore possibilities
for inner shell complexes that  cover the coordination  number range
observed in the simulation were not found.  
Evidently, interactions with the outer shell material can be decisive in
stabilizing coordination structures observed in liquid solution and in
clathrate phases.

Although quasi-chemical approaches are formally exact and efficient in
applications to the  hard sphere fluid \cite{ashbaugh:02,PrattLR:Quatst}
and to ion hydration problems \cite{RempeSB:hydnL+,RempeSB:hydnN+}, these {\em ab
initio\/} quasi-chemical theories applied to Kr(aq) will be more
difficult.  Several contributions have to be assembled that are an
order-of-magnitude larger than the net result for the hydration free
energy.  Modest errors in any one of those several contributions can
easily destroy any reasonable correspondence with experimental results.

We have developed a response to the ``what are we to tell students''
question that is based upon the  known thermodynamic peculiarities of
liquid water and doesn't require specific molecular structural
hypotheses. In addition to that simple argument, we can
tell students several other points of  context.   We can tell students that although  the
primitive concepts of hydrophobic effects based upon water-oil fluid
phase separation are reasonable, the structural  molecular
conceptualizations of hydrophobic effects, such as clathrate  models,
have achieved {\em no\/} consensus.  A lack of
specificity and quantitativeness  in the thermodynamic analyses 
of these molecular pictures contributes  to this lack of consensus.  We  can tell students
that there is  an orthodox clathrate picture that is
not literally correct.  A reformed `clathrate-like' picture
 also appears in the extant literature; this reformed picture suggests connections to the
orthodox picture, and could be
regarded as a poetic `explanation,' borrowing Stillinger's identification
of a limiting possibility. Established  quantitative connections to signature thermodynamic properties aren't available for either the orthodox or the reformed  `clathrate-like' pictures
of hydrophobic  hydration.  
	
Despite the fact that the various molecular  scale pictures have not
been proven, a role for molecular theory in this setting can follow the
conventional understanding of scientific inference: approximate theories
should provide clear expression of a physical idea at a molecular level
and should be quantitatively testable.  The degree of success in those
tests then  contributes to the degree of confidence in the physical
pictures. Molecular mechanisms that are tested only at an
impressionistic level, even after an  investment in quantitative
computer simulations, are less conclusive.

This discussion leads to consideration of Ockham's razor
\cite{ockhamsrazor}.  For thermodynamics  of primitive hydrophobic
effects, we have simple, logical, well-tested molecular  theories \cite{PrattLR:Molthe} and
those theories don't explicitly involve structural mechanisms such as
clathrate models.  Additional hypotheses  in response to ``what are we
to  tell students'' aren't required at this stage.

\section*{Acknowledgements}
The work was supported by the US Department of Energy, contract
W-7405-ENG-36, under the LDRD program at Los Alamos and Sandia. LA-UR-02-6362.

\pagebreak


\end{document}